\documentclass[12pt,preprint]{aastex}

\newcommand{\HI}{\ion{H}{1} }
\newcommand\ST{\rule[-5em]{0pt}{5em}}

\shortauthors{Fern\'andez et al.}

\usepackage{graphicx}
\usepackage{subfigure}
\usepackage{epsf}
\usepackage{longtable}
\usepackage{lscape}
\usepackage{rotating}

\begin{document}

\title{A Radio Perspective on the Wet Merger Remnant NGC 34}

\author{Ximena Fern\'andez}
\affil{Department of Astronomy, Columbia University, 550 West 120th Street, New York, NY 10027}
\email{ximena@astro.columbia.edu}

\author{J.H. van Gorkom}
\affil{Department of Astronomy, Columbia University, 550 West 120th Street, New York, NY 10027}
\email{jvangork@astro.columbia.edu}

\author{Fran\c{c}ois Schweizer}
\affil{Carnegie Observatories, 813 Santa Barbara Street, Pasadena, CA 91101}
\email{schweizer@obs.carnegiescience.edu} 

\author{Joshua E. Barnes}
\affil{Institute of Astronomy, University of Hawaii, 2680 Woodlawn Drive, Honolulu, HI 96822}
\email{barnes@IfA.Hawaii.Edu}

\begin{abstract}
We present VLA observations of the neutral hydrogen and radio continuum of NGC 34 (= NGC 17 = Mrk 938). This object is an ideal candidate to study the fate of gas in mergers, since, as shown by an optical study done by \citet{SS07}, it is a gas-rich (``wet'') merger remnant of two disk galaxies of unequal mass hosting a strong central starburst and a weak AGN. We detect HI emission from both tidal tails and from nearby galaxies, suggesting that NGC 34 is actually part of a gas-rich group and might have recently interacted with one of its companions. The kinematics of the gas suggests this remnant is forming an outer disk of neutral hydrogen from the gas of the northern tail. We also detect broad HI absorption ($514 \pm 21$ km s$^{-1}$ wide) at both negative and positive velocities with respect to the systemic velocity. This absorption could be explained by the motions of the tidal tails or by the presence of a circumnuclear disk. In addition, we present radio-continuum images that show both nuclear ($62.4 \pm 0.3$ mJy) and extra-nuclear emission ($26.5 \pm 3.0$ mJy). The extra-nuclear component is very diffuse and in the shape of two radio lobes, spanning 390 kpc overall. This emission could be a signature of an AGN that has turned off, or it could originate from a starburst-driven superwind.  We discuss the possible scenarios that explain our observations, and what they tell us about the location of the gas and the future evolution of NGC 34.
\end{abstract}

\section{Introduction}
In the hierarchical galaxy-formation model, mergers play a fundamental role in galaxy evolution \citep{WR78}.  Mergers have been extensively studied observationally and numerically since the pioneering efforts of \citet[e.g.,][]{Z53,Z56} and \citet{TT72}.  It is now widely accepted that a fraction of early-type galaxies have formed from the merging of galaxies \citep[e.g.,][]{S98}.  Many of these mergers are known in the literature as ``wet'' since their progenitor galaxies were rich in atomic and molecular hydrogen. It is therefore essential to understand how these systems may evolve into the gas-deficient early-type galaxies we see today.  A complication is that many of these wet mergers harbor starbursts and/or active galactic nuclei (AGN).  Merger-induced starbursts are beginning to be comprehended, but the processes associated with the formation and growth of black holes are less well understood.  In addition to this, the gas involved in these two phenomena plays a competing role, thus further complicating our understanding of the physics. Both starbursts and AGN use hydrogen as their fuel, requiring that outer gas be sent to the inner regions, but at the same time galactic winds associated with both phenomena provide a mechanism for the gas to escape the central regions of the remnant.  In this study we are interested in understanding the fate of gas in a wet merger, especially in the presence of both a starburst and an AGN, to gain insight into the formation and evolution of early-type galaxies. 

NGC 34 (= NGC 17 = Mrk 938) is an ideal candidate for study because it is a gas-rich merger that hosts a strong starburst and a weak AGN, as evidenced by its optical, infrared, radio and X-ray properties. Throughout this paper we adopt a distance of 85.2 Mpc \citep[][hereafter SS07]{SS07}.  All of the properties we cite have been corrected to this distance. SS07 did a thorough optical study to find detailed clues about the merger process from an optical perspective. Figure 1 is a \textit{B}-band image from their study. The two linear tidal tails to the northeast and south of the main body reveal this to be a remnant of two merged disk galaxies.  In addition to these tails, there are other optical features typical of a merger remnant: a single nucleus, and an envelope that contains dust lanes, ripples, fans, and jets of luminous matter.  Also, there is a cloud of luminous debris to the northwest of the nucleus (NW cloud), whose origin could be either remnant material from one of the progenitors or a companion.  Single-dish radio data show NGC 34 to be gas-rich in both neutral and molecular hydrogen, since the  neutral-hydrogen mass is $M_{\rm HI}\approx 1.7 \times 10^{10} \: M_\odot$ \citep{MW84}, and the  molecular-gas mass derived from CO observations is $M_{\rm H_2}\approx 7 \pm 3 \times 10^9 \: M_\odot$ \citep{K03,K90,C92}. 

NGC 34 has also been studied extensively in the infrared, since it is a luminous infrared galaxy (LIRG).  LIRGs are defined as having infrared luminosities in the range $11.0 < \log(L_{IR}/L_{\odot}) < 12.0$ \citep{S87}. Most LIRGs are interacting/merging systems, with vast quantities of molecular gas ($\sim 10^{10}~M_{\odot}$). At the lower end of this luminosity range, the bulk of the infrared luminosity is due to dust heating by a nuclear starburst, while AGN become increasingly important at higher luminosities \citep{SM96}. The infrared luminosity of NGC 34 is $\log(L_{IR}/L_{\odot}) = 11.61$ \citep{C92}, suggesting that both the starburst and the AGN contribute significantly to it.  Inferred star formation rates range from 50 $M_\odot$ yr$^{-1}$ \citep{V05} to 80--90 $M_\odot$ yr$^{-1}$ \citep{P04}. 

The optical nuclear spectrum of NGC 34 also shows signs of the presence of an AGN and starburst.  It appears composite and exhibits a weak [OIII] $\lambda 5007$ emission line relative to H$\beta$ or H$\alpha$, placing NGC 34 in a transition category of objects with nuclear spectra between starburst and Seyfert 2 \citep[e.g.][]{G99}. Estimates for the relative contributions of the starburst and AGN to the bolometric flux range between 75\%/25\% \citep{IAH04} and 90\%/10\% \citep{G99}.  The X-ray luminosity of NGC 34 serves as independent evidence for the presence of an AGN: $L_{X, 2-10 \: \rm keV} \approx 2.2^{+2.8}_{-0.9} \times 10^{42}$ ergs s$^{-1}$ measured from \textit{XXM-Newton}  observations by \citet{G05} (with errors computed by J.~Rigby, see SS07). This value is too high to be explained by a starburst alone and is comparable to that of the classical Seyfert 1.5 galaxy NGC 4151. 

The optical study by SS07 discovered various properties that give clues about the merging history of NGC 34.  The authors propose the following: two disk galaxies of unequal mass merged, creating a galaxy-wide starburst that occurred first about 600 Myr ago, peaking over 100 Myr ago, and giving birth to an extensive system of young globular clusters with ages in the range 0.1--1.0 Gyr.  The study reveals a young, blue stellar exponential disk that formed about 400 Myr ago from gas settling towards the end of the merger. The two nuclei are either still coalescing or have just finished merging, creating the final concentrated and obscured starburst and the AGN, both driving a strong gaseous outflow.  The authors detect this outflow from the blueshifted \ion{Na}{1} D doublet, which yields a mean outflow velocity of $-620 \pm 60 \: \rm km \: s^{-1}$ and a maximum velocity of $-1050 \pm 30 \: \rm km \: s^{-1}$, and suggest that the outflow may extend northward of the nucleus in a fanlike structure.  These velocities are surprisingly high, even for a galaxy with a high SFR such as NGC 34, indicating that the outflow could be linked to AGN activity. 

In the following sections, we present a comprehensive radio study of NGC 34 that includes \HI and radio continuum observations to analyze the distribution and kinematics of the gas.  We seek to understand the fate of the gas, which is already exhibiting intriguing behavior in the optical study as evidenced by the detection of the blue stellar disk formed in the later stages of the merger and the ouflow of cool gas.  Our observations allow us to address the following questions:  Were the two progenitors gas-rich? Where is the neutral gas located?  Is there any neutral gas left in the center of the remnant?   Is the \HI feeding the starburst and/or AGN, or is it being pushed out of the central regions via superwinds?  What is the origin of the NW cloud?  In addition to answering these questions, the velocity information of the tails will help put tight constraints on the modeling of unequal mass mergers and accurately determine the mass ratio of the progenitors, the merger history, and the evolution of the remnant.

\section{Observations and Data Reduction}
NGC 34 was observed in 2008 and 2009 with the Very Large Array (VLA)\footnote {The VLA is operated by the National Radio Astronomy Observatory, which is a facility of the National Science Foundation (NSF), operated under cooperative agreement by Associated Universities, Inc.} in spectral-line mode at 21-cm and in two hybrid configurations: the DnC array and CnB array \citep{NTE83}. These hybrid arrays consist of the antennas in the east and west arms being in the more compact configuration (first letter), while the ones in the north arm are in the more extended configuration (second capital letter).  At the adopted distance of 85.2 Mpc for NGC 34 (SS07), these hybrid configurations give a resolution of 18 kpc and 6 kpc, respectively.  The DnC array observations of 2008 consisted of one run lasting 5 hours, while the CnB array observations of 2009 consisted of 3 runs, each lasting 7 hours. Both sets of observations used a 6.25 MHz bandwidth centered at the heliocentric systemic velocity of 5870 km s$^{-1}$ (SS07), with 31 channels and a spacing of 43 km s$^{-1}$. This setup resulted in a velocity coverage of 1285 km s$^{-1}$, starting at 5229 km s$^{-1}$, and ending at 6514 km s$^{-1}$.

The data were reduced with the Astronomical Image Processing System (AIPS) following standard calibration procedures for each configuration.  We subtracted the continuum by making a linear fit to the line-free channels, and then combined the data of the two configurations in the visibility plane.  We generated three HI data cubes, one for each configuration and one for the combined data. The resulting data cubes were made with a robustness parameter of 1 \citep{B95}, a combination between natural and uniform weighing, to optimize both sensitivity and spatial resolution. We CLEANed the data cubes to minimize sidelobes.  Maps of the total HI distribution and velocity field for the combined observations were made by taking moments along the frequency axis. This was done by first smoothing the data cubes spatially with a Gaussian function over a cellsize of 15\arcsec\ and in velocity with a Hanning function over a cellsize of 3 channels. A mask was created by blanking pixels of the smoothed data cubes below our cutoff set at 2$\sigma$. This mask was applied to the full-resolution data cube to then sum over the data and generate the different moment maps.

Additionally, we present untapered and tapered images of the radio-continuum that were generated by combining the line-free channels in the visibility plane. We made the untapered image with a robustness parameter of 1 and CLEANed it by placing boxes around the emission regions.  We also generated a tapered image with the same parameters, but with point sources subtracted from the uv data, and applying a Gaussian taper of 5 k$\lambda$ to better map the extended structure. 

Table 1 presents information about the radio observations for each configuration, the combined data set, and the continuum images.  It lists the number of channels, calibrators used, synthesized beam size, rms noise level, conversion factor from flux (mJy) to brightness temperature (K), and the \HI column density sensitivity (corresponding to $1\sigma$).

We calculated \HI masses via the following equation:
\begin{equation}
M_{\rm HI} (M_ {\odot})= 2.36 \times 10^{5}~D_{\rm L}^{2} \int{S dv}
\end{equation}
where $\int{S dv}$ is the integrated flux density in Jy~km~s$^{-1}$, and $D_{\rm L}$ is the luminosity distance in Mpc.

We utilized the following equation to calculate \HI column densities (assuming the gas is optically thin):
\begin{equation}
N_{\rm HI} (\rm cm^{-2}) = 1.823 \times 10^{18} ~T_b~ \Delta \it{v}
\end{equation}
where $T_b$ is the brightness temperature in K, and $\Delta v$ is the velocity width in km s$^{-1}$.  We determined \HI-absorption column densities via the following equation:   
\begin{equation}
N_{\rm HI}  (\rm cm^{-2}) = 1.823 \times 10^{18}~ T_{\rm s}~ \int{\tau\it{dv}}
\end{equation}
where $T_{\rm s}$ is the spin temperature, and $\int{\tau dv}$ is the integrated optical depth.

\section{Results and Analysis}

\subsection{\HI Observations}

\subsubsection{Channel Maps}
Figure 2 shows a set of individual channel maps of our combined observations overlaid on a deep image from SS07, with contour levels of ($-$1.04, $-$0.78, $-$0.52, $-$0.26, 0.26, 0.52, 0.78, 1.04) mJy/beam.  The optical heliocentric systemic velocity of  NGC 34 is $5870 \pm 15$ km s$^{-1}$ (SS07).  We detect emission from three kinds of features: the tidal tails, emission south of the nucleus, and neighboring galaxies. The \HI coincident with the optical northern tail is seen in panels 5--10 and with the southern tail in panels 4--6.  In addition to this, there is emission south of the nucleus  wrapping around the remnant and reaching the NW cloud in panels 3--10.  The kinematics of the tails suggests this gas may actually be part of the northern tail that has fallen back towards the nucleus and is forming an outer disk of \HI.  The tip of the northern tail is moving at redshifted velocities (panels 8--10), while the emission closest to the nucleus is moving at blueshifted velocities (panels 5--7), showing that part of the tail experiences a reversal of velocities and is approaching the nucleus.  This emission is continuous with the HI seen to the south of the nucleus, suggesting they could be connected. This gas seems to be rotating as it wraps around the nucleus reaching the NW cloud at redshifted velocities.  We detect broad absorption against the continuum source that begins at the blueshifted velocity of 5570 km s$^{-1}$ and ends at the redshifted velocity of 6084 km s$^{-1}$, spanning over 500 km s$^{-1}$.

In addition to the hydrogen detected in NGC 34, we see \HI emission from four companion galaxies, showing that NGC 34 belongs to a small gas-rich group (see Fig.~5 below).  None of the companions has previously been observed in \HI. Out of these, NGC 35 is the biggest one and lies at a projected distance of 131 kpc from NGC 34.  The other three are smaller galaxies, of which two had been catalogued by \citet{P03} and one is uncatalogued.   Table 2 gives the designated names of the companions, their coordinates, their systemic heliocentric \HI velocity, their flux density, and the corresponding \HI mass. We calculate the \HI mass for NGC 34 and neighboring galaxies from the channel maps using equation 1 (assuming the adopted distance of 85.2 Mpc). We get a total \HI mass of ($7.2 \pm 0.2) \times 10^9 M_\odot$ for NGC 34. We note this is a lower limit since there is absorption.  This value is lower than the published value of $1.7 \times 10^{10} \: M_\odot$ \citep{MW84} from single-dish observations.  The single-dish mass is higher since NGC 35 (which is 5\arcmin\ away) was also in the 8.2\arcmin\ beam.  In addition to this, the value is uncertain due to errors in baseline fitting.  We therefore consider our value to be consistent with the single-dish data.

\subsubsection{Position--Velocity Diagram}
Figure 3 shows a position--velocity diagram along the northern tidal tail and through the nucleus to the gas southwest of it.  This diagram was made by rotating the data cube by $38\fdg4$, which corresponds to the position angle along the northern tidal tail, around the velocity axis going through the nucleus.  We then transposed the axes such that the velocity and declination were the first two axes, leaving right ascension as the third one.  We plot a 15\arcsec\ wide slice of this transposed cube at the central pixel to show the kinematics of the northern tail and of the gas southwest of the nucleus at different declinations. The dashed lines correspond to \HI absorption, while the solid lines show the \HI emission.  Note that the velocities in the northern tail start redshifted and become blueshifted as they move towards the nucleus, while the velocities in the southwestern gas are mostly blueshifted.  Also, the central \HI absorption is much wider in velocity than the \HI emission from the northern tail and southwestern gas. 

\subsubsection{Absorption Profile}
Figure 4 shows the absorption profile of our combined observations at the central position of the remnant. The profile is $514 \pm 21$  km s$^{-1}$ wide and asymmetric with respect to the adopted systemic velocity of 5870 km s$^{-1}$. The blueshifted component has a width of 300 km s$^{-1}$ and reaches its maximum depth at 5613 km s$^{-1}$, while the  redshifted component has a width of 214 km s$^{-1}$ and reaches its maximum depth at 5956 km s$^{-1}$.  We calculate a column density of $3 \times 10^{21}$ cm$^{-2}$ using equation 3, where the integrated optical depth is 16.4 km s$^{-1}$, and where we assume a value of 100 K for the spin temperature.

\subsubsection{Moment Maps}
Figure 5 shows the moment 0 map, which is the total \HI distribution of our combined observations overlaid on a \textit{Digitized Sky Survey} image (top) and on a deep image from SS07 (bottom). The contours represent emission drawn at levels of (8, 28, 48, 68, 108)$\,\times\,10^{19}$ cm$^{-2}$. 
We detect emission from both tidal tails, an outer disk wrapping around the nucleus, and from neighboring galaxies.  This image also suggests that the NW cloud and northern tail may be a single feature, since we see continuity in the projected \HI distribution.  The hydrogen in the northern tail reaches a projected distance of 52 kpc from the center and is significantly displaced to the east when compared to the stars. The southern tail has a small amount of \HI coincident with the stars, reaching a maximum projected distance of 49 kpc from the center.  The gas in this tail is also significantly displaced to the west when compared to the optical. Some of the gas south of the nucleus might also contain \HI from the southern tail.

Figure 6 shows the moment 1 map color coded by velocity. This image of the velocity field of NGC 34 helps visualize the kinematics of the \HI gas. As already evidenced by the channel maps and position--velocity diagram, the map shows that the velocity field along the tails is surprisingly smooth.  We can see the tip of the northern tail at redshifted velocities, while the \HI near the nucleus is moving at blueshifted velocities.  This gas fits in with the velocity pattern south of the nucleus where we see the \HI rotating since we detect the gas at blueshifted and then redshifted velocities.

\subsection{Radio Continuum}
To better understand the central \HI absorption we also made images of the radio continuum. These images portray a complex picture of the radio morphology of NGC 34.  Besides two point sources, of which one coincides with NGC 34, they reveal two extended structures to the northeast and west  that resemble faint radio lobes. Figure 7 shows the radio continuum contours drawn in levels of (0.2, 0.4, 0.6, 0.8, 1) mJy/beam from the untapered image and overlaid on a \textit{Digitized Sky Survey} image. These contours show the central emission centered at RA~= 00:11:06.558 $\pm$ 0.011, Dec~= $-$12:06:27.50 $\pm$ 0.14, surrounded by hints of extended structure, and emission coming from NGC 35 to the north. Figure 8 provides a closer look at the nuclear and extra-nuclear continuum emission.  The top panel depicts the strong nuclear emission with contours drawn at levels of (2, 4, 6, 8, 10, 20, 40) mJy/beam, overlaid on a deep image of SS07. The radio continuum shows an unresolved point source of $62.4 \pm 0.3$ mJy coincident with the optical nucleus, accompanied by a fainter point source 41\arcsec\ (17 kpc) to the south with no optical counterpart.  This fainter source of $5.1 \pm 0.1$ mJy is probably a background object, while the stronger source is emission from the AGN or from the highly concentrated starburst in NGC 34.   We subtracted the point sources and made a tapered image to better map the extra-nuclear emission.  The bottom panel of Figure 8 is a false-color image of the diffuse continuum, showing two faint radio lobes or bubbles.  These structures extend $15\farcm7$ (390 kpc) in projection, and have a total radio flux of $26.5 \pm 3.0$  mJy.  We confirmed the validity of these radio lobes with a D-array EVLA observation of 2 hours made on 2010 Aug 7.

\section{Discussion}

\subsection{Location and Kinematics of the Neutral Gas}
Our \HI observations show a complicated picture of the location of the gas in NGC 34 as evidenced by the neutral hydrogen detected in emission and absorption.  We detect emission from the two tidal tails, indicating both progenitors were gas-rich. We expected to detect hydrogen emission from the northern tail since there are OB associations and ionized gas in it (SS07).  The \HI in the southern tail reveals that the less massive progenitor had less gas than the northern tail.  The position--velocity diagram (Figure 3) provides an overview of the kinematics of the neutral gas.  It shows that the gas seen in absorption has a wider velocity range than the gas seen in emission along the northern tail and southwestern gas.  The optical outflow detected by SS07 shows an even wider velocity range than the \HI in absorption, but it is important to note that our observations did not cover that full velocity range. The position--velocity diagram also shows smooth velocity gradients along the northern tail and southwestern gas, indicating the hydrogen is moving in regular patterns. As discussed earlier, the NW cloud may be a continuation of the northern tail, which therefore might indicate that some of the gas in the tail is falling back towards the nucleus and wraps around the remnant. This could be evidence for gas settling into an outer ring, which would be consistent with the exponential young stellar disk studied by SS07.  Since it is believed that young disks form from the inside out, the inner regions of the disk may already have formed stars (as evidenced by SS07), and the NW cloud could be outer layers of gas settling. The neutral gas in both tails is noticeably displaced when compared to the stars, indicating there might be a mechanism pushing the gas aside.

Another goal of this study is to determine whether there is neutral hydrogen left in the center, since this could fuel the starburst and/or the AGN.  Our observations can place a limit on the proximity of the neutral gas to the center of the remnant by comparing the two absorption profiles for the CnB array and DnC array (Figure 9). Due to the fact that our beams are  larger than the continuum source,
the absorption profiles taken at different resolution could be a combination of \HI emission and absorption.  If there were \HI emission throughout the central regions, we would have detected a deeper absorption feature when observing with a higher resolution.  As the figure shows, the absorption depth is nearly the same at blueshifted velocity, and there is some emission centered around the systemic velocity in the absorption profile of the DnC observations that is not seen in the higher resolution data. This suggests there is some emission within 18 kpc at about the systemic velocity, with a peak of 1.5 mJy/beam at around 5700 km s$^{-1}$ and an approximate width of 400 km s$^{-1}$.  The peak of this emission corresponds to a column density of $6.2 \times 10^{19}$ cm$^{-2}$.  We can also estimate the central \HI mass by summing the differences between the two absorption profiles to get a flux density, and using equation 1.  We find $7.3 \times 10^8 M_\odot$ of \HI in the inner 18 kpc of NGC 34. Hence,  we see evidence for \HI emission at about the systemic velocity on scales larger than 6 kpc, but smaller than 18 kpc.   When making this point, however, it is important to recall that there are also substantial quantities of molecular hydrogen in this remnant, whose spatial distribution we do not know in detail. If the H$_{2}$ is confined to the inner 18 kpc of the remnant and there is $7.3 \times 10^8 M_\odot$ of \HI, this would translate to a mass ratio of H$_{2}$ to \HI of 10:1.  The distribution of the gas in the center of NGC 34 might then be similar to that in NGC 7252 \citep{H94}, where most of the \HI is found in the outer regions, while the molecular gas is confined to the center \citep{W92}.  This would imply that the \HI has been transformed into H$_{2}$, providing fuel for star formation to take place.  It is therefore important to map the molecular gas distribution within NGC 34 to better understand the gas in different phases. 

In addition to detecting \HI in four companion galaxies, we also detect a hint of emission between NGC 34 and its biggest companion NGC 35.  This is best seen in our DnC array observations at the systemic velocity (Figure 10).  The amount of \HI seen between the two galaxies (excluding the material of the northern tail) corresponds to a mass of about $2.5 \times 10^8 M_\odot$. If this emission is real, it may suggest that the two galaxies had a relatively close encounter in the recent past.  
 
\subsection{Puzzling absorption}
The broad blueshifted and redshifted absorption of the atomic gas in NGC 34 is a striking signature that calls for an explanation.  Detecting this absorption profile means that there is hydrogen moving toward us and away from us along our line of sight to the nucleus. We suggest two possible scenarios to explain this behavior, depending on whether the \HI gas has settled or not.  

In the first scenario, the absorption may be explained in terms of the motions of the tidal tails or of non-circular orbits of unsettled gas.  We can examine the kinematics of the tidally extracted gas more closely by analyzing ionized-gas velocities measured at higher spatial resolution along the northern tail from an optical spectrum obtained with the 6.5 m Clay telescope at Las Campanas (Schweizer et al., in preparation). Figure 11 shows the \HI position--velocity diagram of Figure 3 overlaid with these ionized-gas velocities (data points), measured also at a position angle of $38\fdg4$ and in steps of $1\farcs13$ along the $1\farcs0$ wide slit.  The northern tail contains ionized gas (SS07), which is clearly seen in this diagram at both blue- and redshifted velocities relative to the nucleus.  Except near the tip of the optical northern tail, toward the center the ionized gas moves at negative  relative velocities, eventually reaching the systemic velocity and becoming redshifted southwest of the nucleus.  This suggests  that the northern tail experiences a reversal in velocities, approaches the central regions, and then wraps around the remnant. We could be seeing the blueshifted and redshifted absorption as the \HI passes in front of the strong continuum source. This is consistent with the \HI emission but does not fully explain the high-velocity range we see in the absorption.  

These high velocities suggest there is \HI moving at smaller distances from the nucleus, thus motivating us to propose a second scenario.  In this scenario, the gas may have settled into a circumnuclear disk.  \citet{M08} observe a similar absorption profile in Centaurus A, where they detect a broad (400 km s$^{-1}$) line asymmetric with respect to the systemic velocity.  The authors interpret this profile as evidence for a circumnuclear gas disk since the distribution and kinematics of the molecular gas in that galaxy resemble those of the neutral hydrogen.  Circumnuclear disks often have a mix of molecular and neutral hydrogen. Thus, it is essential to map the molecular gas content of NGC 34 to determine whether it coincides with the \HI or not.  We know from single-dish observations by \citet{C92} that the velocity width of the CO(1--0) line is about 500 km s$^{-1}$, starting at approximately 5500 km s$^{-1}$ and ending at 6000 km s$^{-1}$, thus matching our measured \HI absorption-line width of 514 km s$^{-1}$. We now need to know whether the distribution of the molecular gas traces the \HI absorption spatially and in velocity, so follow-up observations  are essential.  If this were to be the case, we would have convincing evidence for the presence of a circumnuclear disk.

\subsection{Radio Continuum}
We detect both nuclear and extra-nuclear radio-continuum emission in NGC 34.  Previous radio observations of NGC 34 have shown the nuclear component at different resolutions \citep[e.g.,][]{L93,C98,T00}.  Our observations show an unresolved point source coincident with the optical nucleus. An upper limit for the FWHM of the central point source is 6 kpc (15\arcsec), which is the resolution of our CnB observations.  We see other point sources present in the radio continuum image (Figure 7) as well, but those are either associated with other galaxies (e.g., NGC 35) or with likely background quasars. 

The origin of the nuclear emission (62.4 mJy) in NGC 34 can be further explored by examining the higher-resolution data available in the literature. NGC 34 yielded only a single baseline detection when observed with VLBI \citep{L93}, showing that the nucleus of this remnant is barely detectable on VLBI scales.  This suggests that the nuclear radio-continuum emission is extended and mostly due to the highly concentrated starburst. Observations by \citet{T00} with the VLA A array at 8.4-GHz show a slightly resolved source of 14.5 mJy and 149 pc ($0.4^{\prime\prime}$) in size.  Once more, this suggests the radio-continuum emission is mostly due to the starburst. As an additional check, we calculate the SFR from the radio luminosity and compare it to the SFR determined from the IR luminosity.  We use the following relation derived by \citet{A03}: 

\begin{equation}
\mathrm{SFR}_{1.4\mathrm{GHz}}=\frac{L_{1.4\mathrm{GHz}}}{8.4 \times 10^{20}~ \rm \mathrm{W~Hz}^{-1}}~M_\odot~\mathrm{yr}^{-1}   
\end{equation}

We get a value of 64 $M_\odot$ yr$^{-1}$, which is consistent with the SFR derived from IR luminosities (see \S 1).  
  
The extra-nuclear radio-continuum emission is very faint and is in the shape of two radio lobes (Fig.\ 8, bottom panel).  We propose two possible explanations for this extra-nuclear emission.  First, we could be seeing fossil lobes created by an AGN that has turned off. This would fit in nicely with what we know about this merger remnant, since we know it has a weak AGN, there is not a lot of atomic gas left in its inner regions, and the gas in the tails is displaced from the optical in the direction of the radio lobes. Fossil lobes could be showing that the AGN was once much more powerful and efficient at clearing out the gas, but that it now has turned off and we detect only a weak signature of the process.

A second possibility is that this could be evidence for a starburst-driven superwind. A study of extra-nuclear diffuse emission in Seyfert galaxies was made by \citet{B94}.  The emission they examined is often very similar to the one seen in NGC 34, two faint radio lobes or bubbles along the minor axis of the host galaxy.  They concluded that this is likely evidence for a starburst-driven superwind and not an AGN, since the emission does not align with the major axis of the nuclear radio emission, suggesting the emission mechanism has to be different.  This would also agree with the properties of NGC 34, since we know there is a strong outflow of cool gas (SS07).  Further optical studies of NGC 34 will be necessary to map the extent of this outflow.  If the outflow were to track the radio emission, there would be a stronger case for a starburst-driven superwind having created the lobes. What is the effect of a superwind on this remnant? A possible outcome is suggested by the fact that the \HI in the northern tail is displaced to the east when compared to the stars.  Superwinds might play an important role in dispersing the gas of wet merger remnants.

\section{Summary and Conclusions}
Our \HI and radio-continuum observations offer insight into the merging history and evolution of NGC 34. Figure 12 compares the extent of the neutral-hydrogen and radio-continuum emissions.  The \HI observations show that NGC 34 has a substantial quantity of gas left and seems unlikely to become an elliptical.  The fact that this remnant hosts both a starburst and an AGN complicates our understanding of the gas, since it is difficult to disentangle the feedback of the two on the gas.  Our data have revealed important aspects related to the radio properties of this remnant:
\begin{itemize}
\item{We detect neutral hydrogen with smooth velocity gradients in both tails, showing the two progenitors were gas-rich. The velocity information will be used to constrain future simulations of the merger remnant to be done with \textit{Identikit} \citep{BH09}. These simulations may enable us to understand the history and evolution of the remnant, and to determine the mass ratio between the two progenitors more accurately.}
\item{We detect the NW cloud in \HI as well, but do not see any sharp distinction between this feature and the northern tail, suggesting that both may be part of the same structure.  This could be evidence for a new outer gas ring/disk that is beginning to form.}
\item{We detect \HI emission from four companion galaxies, showing that NGC 34 is part of a gas-rich group and that it might have interacted with NGC 35.}
\item{Comparing the \HI absorption profiles of observations made with the two array configurations (CnB and DnC) allows us to set limits on the emission near the center of the remnant.  We see evidence for \HI emission at the systemic velocity only on scales larger than 6 kpc and smaller than 18 kpc.}
\item{The blueshifted and redshifted \HI absorption could be evidence of a circumnuclear disk of neutral and molecular gas, or---alternatively---could be related to the tidal tails.  Understanding the nature of this absorption should eventually enable us to determine whether the inner gas has settled or not.  For this, we are planning to map the molecular-hydrogen distribution with CO observations to determine whether there is a circumnuclear gas disk in NGC 34.}
\item{The nuclear radio-continuum emission is mostly extended, indicating that it is dominated
by the central starburst.}
\item{Diffuse, very extended radio-continuum emission in the form of two radio lobes could be evidence for an AGN having recently turned off or for a starburst-driven superwind.}
\end{itemize}

These various aspects lead us to conlude that NGC 34 is {a complicated system where we see the effects of both the starburst and the AGN on the gas.  We see evidence for some of the gas settling into a circumnuclear disk and an outer disk.  We also detect a significant displacement of the gas in both tidal tails, but we do not see direct evidence of depletion.  We can estimate a lower limit for how long it will take NGC 34 to run out of gas.  If we assume there is (1) a mechanism  by which all of the atomic gas returns to the center, (2) a constant star formation rate of 70 $M_\odot$ yr$^{-1}$, and (3) a combined molecular and neutral hydrogen mass of $1.4 \times 10^{10} \: M_\odot$, then it would take the remnant approximately 200 Myr to convert all of its gas into stars. In reality, it may well take significantly longer since most of the \HI is in the tidal tails and the star-formation rate is likely to decrease with time.

\acknowledgments

This research has made use of the NASA/IPAC Extragalactic Database (NED), which is operated by the Jet Propulsion Laboratory, California Institute of Technology, under contract with the National Aeronautics and Space Administration. We also acknowledge use of the HyperLeda database (http://leda.univ-lyon1.fr). M.X.F. gratefully acknowledges the support of the Columbia University Bridge to Ph.D. Program in the Natural
Sciences. This work was supported in part by the National Science Foundation under grant 06-07643 to Columbia University.

\clearpage
\begin{figure}
\begin{center}$
\begin{array}{cc}
\includegraphics[width=2.8in, height=2.8in]{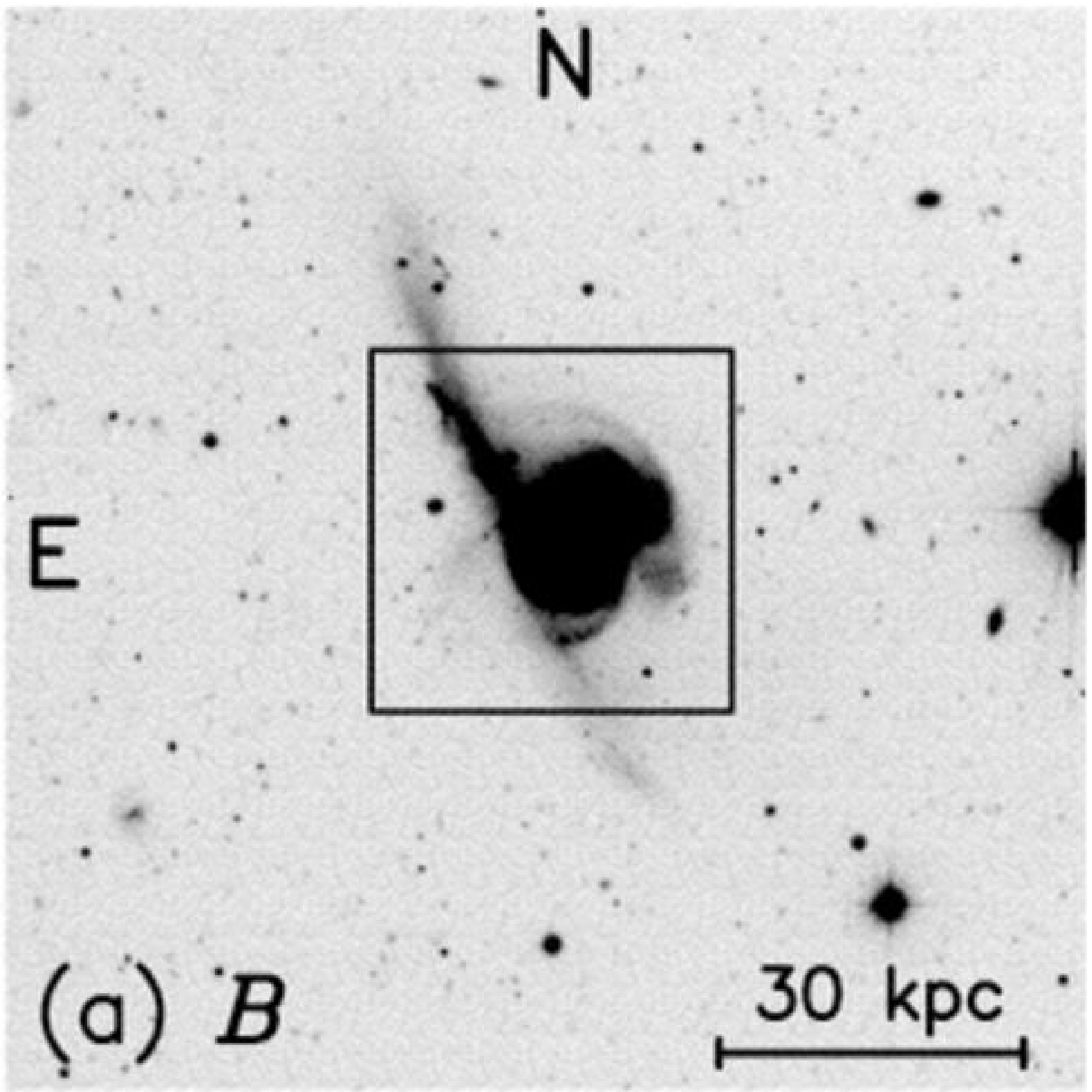}&
\ST \includegraphics[width=2.8in, height=2.8in]{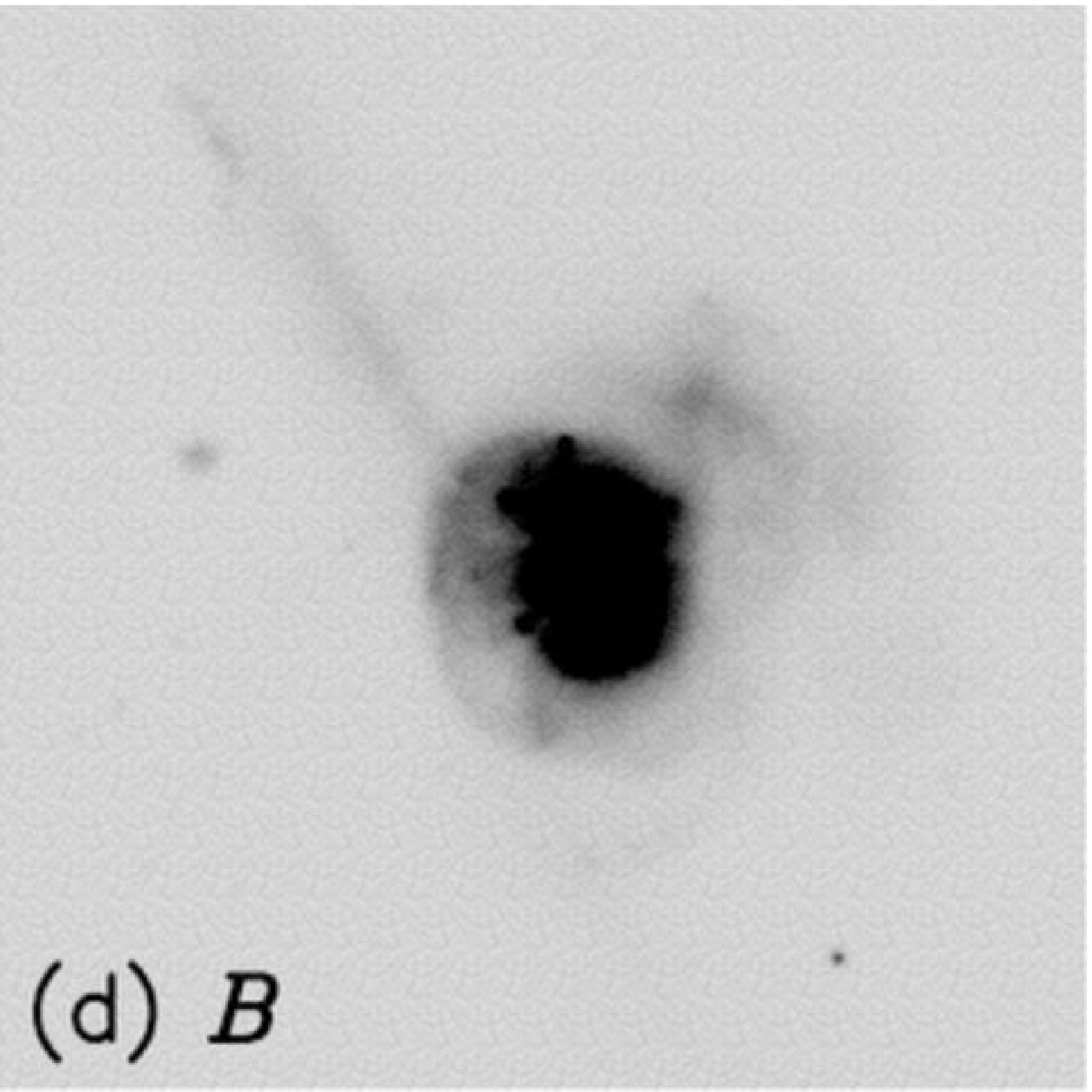}\\
\end{array}$
\caption{\textit{B}-band images of NGC 34 taken with the du Pont 2.5 m telescope.  Images correspond to panels (a) and (d) of Figure 1 of SS07. \textit{Left}: $4\farcm3\times 4\farcm3$ field of view; the box marks the $86\arcsec\times 86\arcsec$ field of view shown enlarged to the right. \textit{Right}: Boxed portion of left image shown enlarged by a factor of 3 and displayed at a lower contrast.}
\end{center}
\end{figure}

\begin{figure}
\begin{center}
\includegraphics[trim = 10mm 5mm 0.5mm 2mm,clip,scale=0.42]{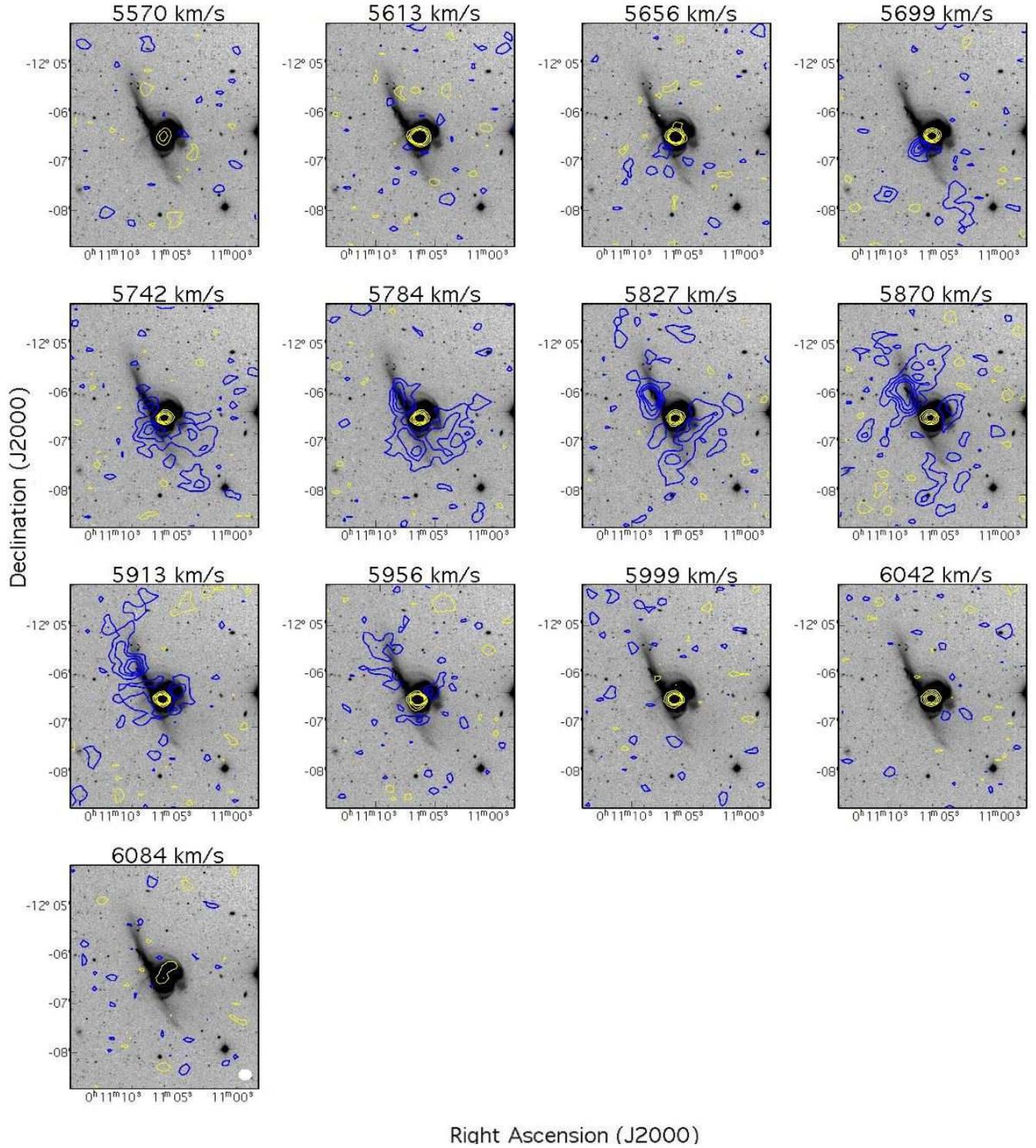}
\caption{Set of individual channel maps of our combined \HI observations of NGC 34 overlaid on a deep image from SS07 with contour levels of ($-$1.04, $-$0.78, $-$0.52, $-$0.26, 0.26, 0.52, 0.78, 1.04) mJy/beam.  Negative contours are marked in yellow, and positive contours in blue. The optical systemic velocity of the remnant is 5870 km s$^{-1}$. The size of the synthesized beam is plotted in the bottom right hand corner of the last panel.  }
\end{center}
\end{figure}

\begin{figure}
\begin{center}
\includegraphics[trim = 20mm 10mm 10mm 10mm,scale=1]{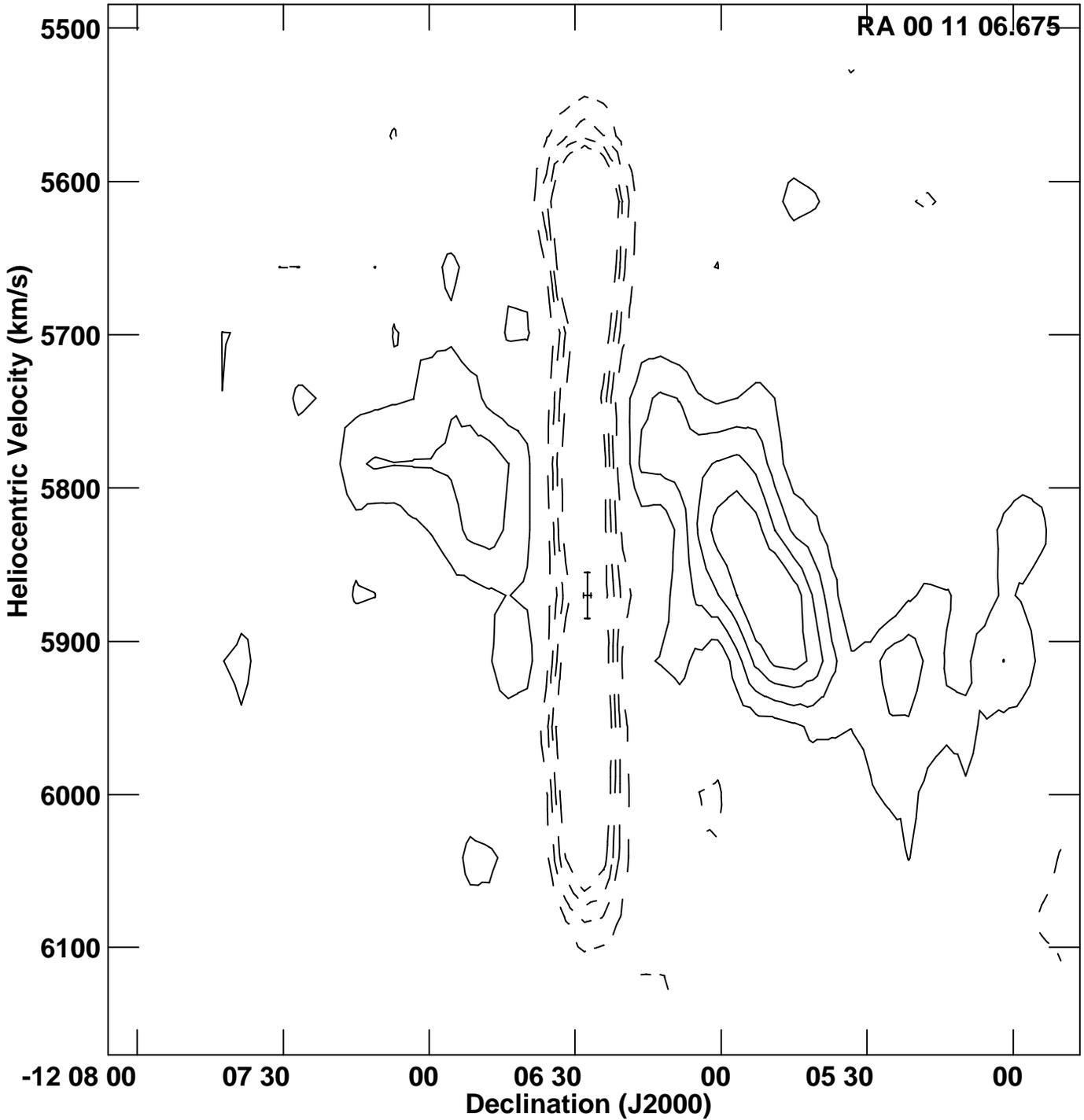}
\caption{Position--velocity diagram along the northern tidal tail (right side) and through the nucleus to the gas southwest of it (left side). The solid contours represent emission in levels of $2\sigma, 4\sigma, 6\sigma, 8\sigma$ while the dashed ones show the absorption in levels of $-2\sigma, -4\sigma, -6\sigma, -8\sigma$. The cross marks the central declination and optical systemic velocity.}
\end{center}
\end{figure}

\begin{figure}
\includegraphics[trim = 10mm 10mm 10mm 10mm,scale=1]{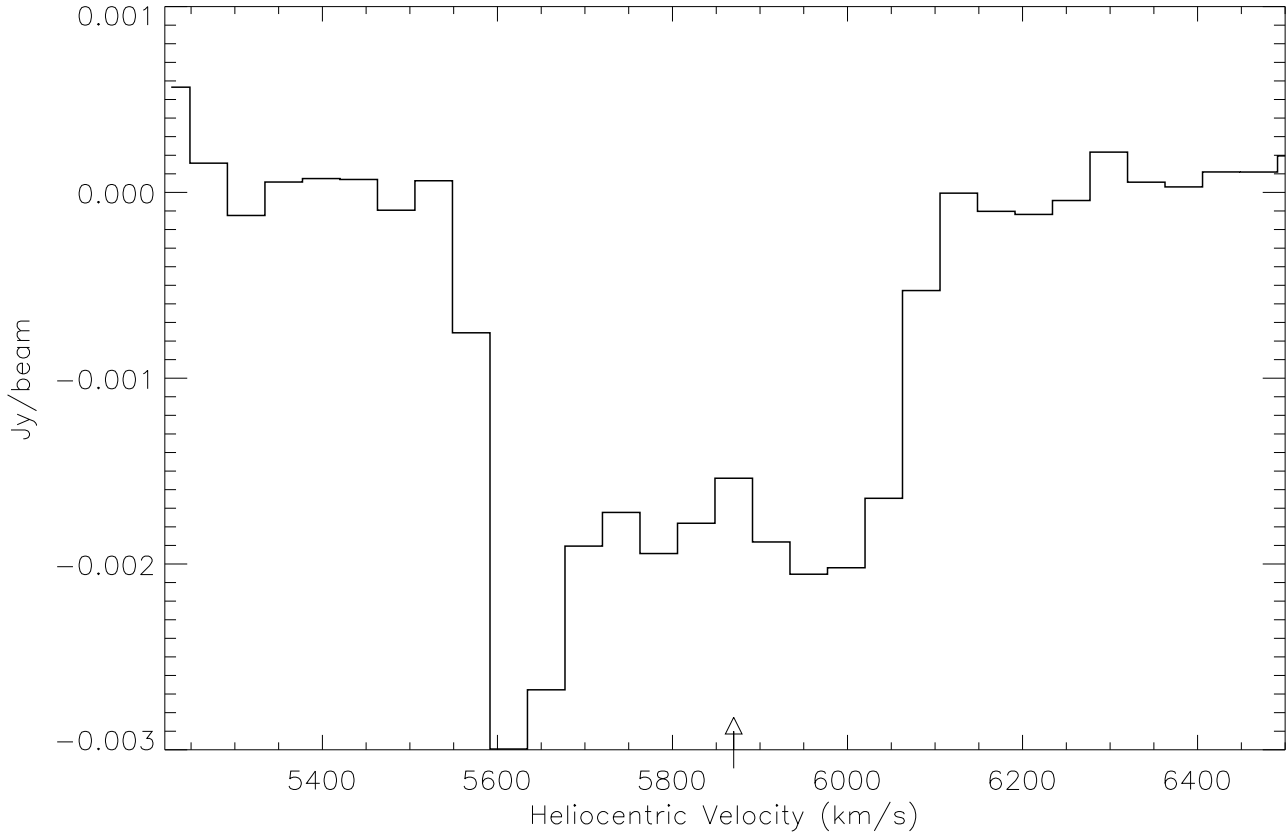}
\vspace{0.5in}
\caption{Absorption profile of our combined \HI observations showing asymmetric blueshifted and redshifted components.  The optical systemic velocity of NGC 34 is 5870 km s$^{-1}$ (indicated by the upward arrow).}
\end{figure}

\begin{figure}
\begin{center}
$\begin{array}{c}
\ST \includegraphics[trim = 10mm 50mm 10mm 40mm,scale=0.9]{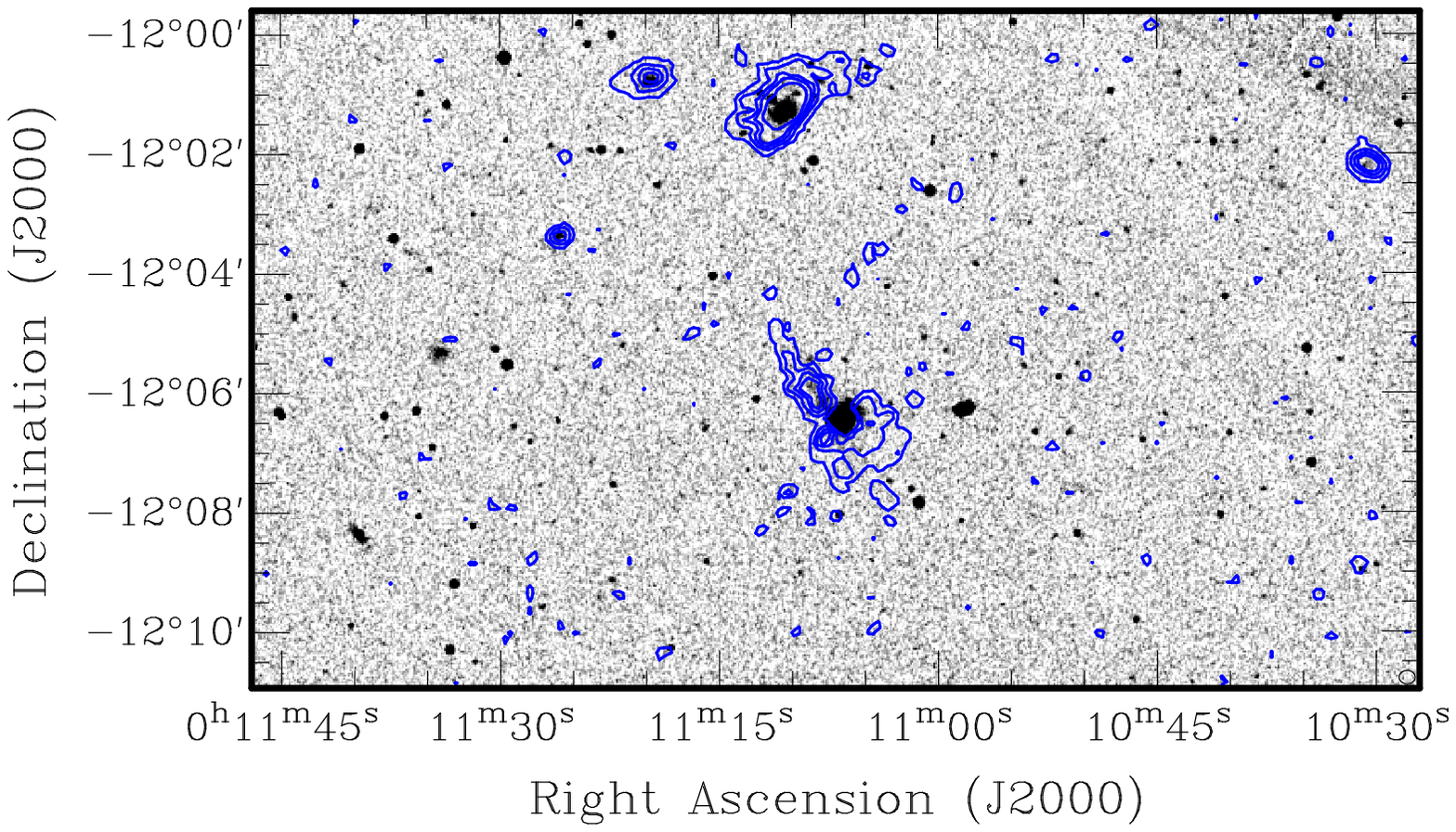}\\
\ST \includegraphics[trim = 10mm 40mm 10mm 40mm,scale=0.8,scale=0.73]{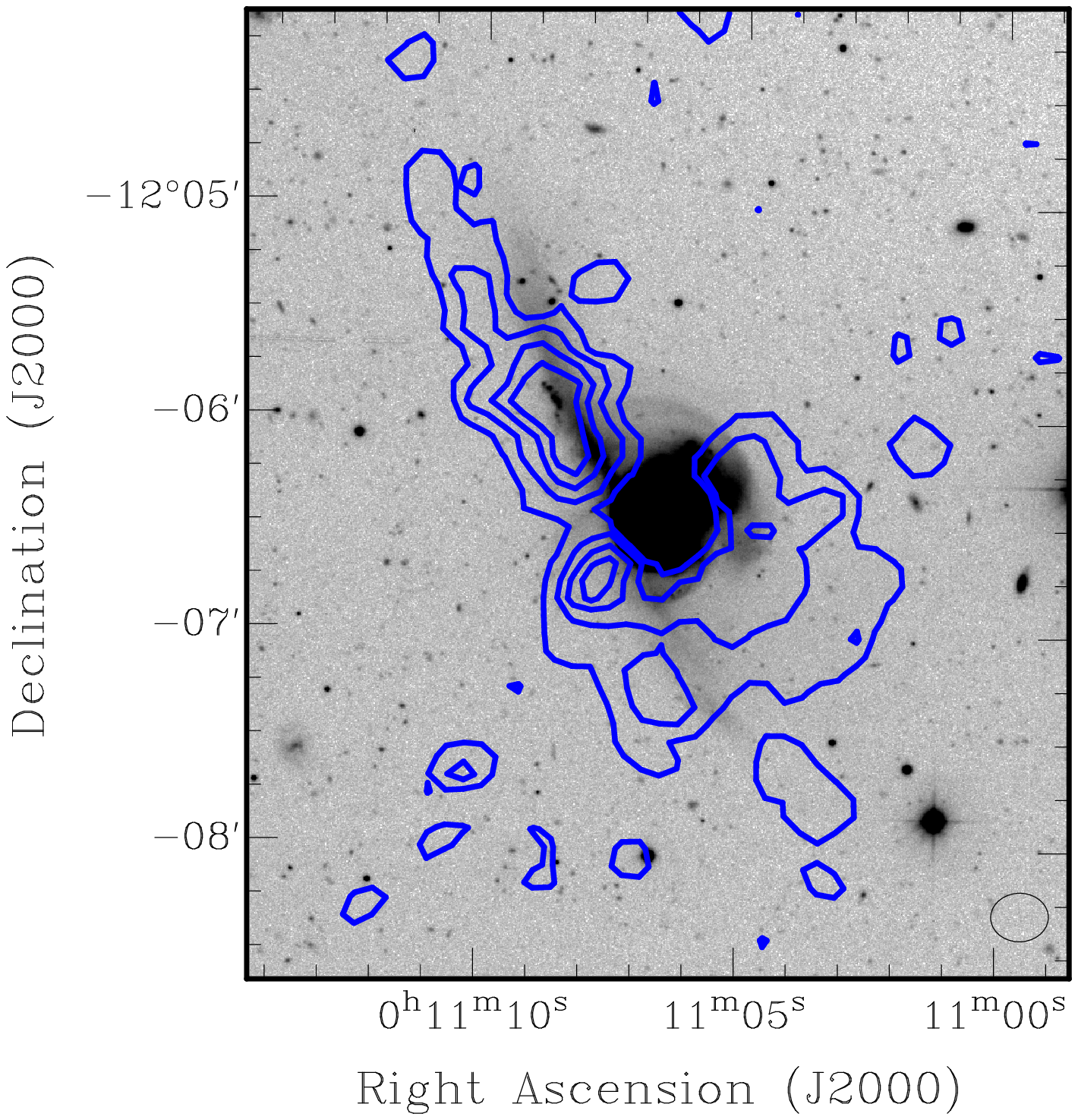}\\
\end{array}$
\vspace{0.3in}
\caption{Moment 0 Map: Total \HI distribution of our combined observations. The blue contours represent \HI emission drawn at levels of (8, 28, 48, 68, 108) $\times 10^{19}$ atoms cm$^{-2}$. The size of the synthesized beam is shown in the bottom right hand corner. \textit{Top:} \HI of NGC 34 and four companions overlaid on a \textit{Digitized Sky Survey} image. \textit{Bottom}: Closer look at the \HI of the tidal tails overlaid on a deep image from SS07.}
\end{center}
\end{figure}

\begin{figure}
\begin{center}
\includegraphics[trim = 1mm 1mm 1mm 1mm,clip,scale=0.95]{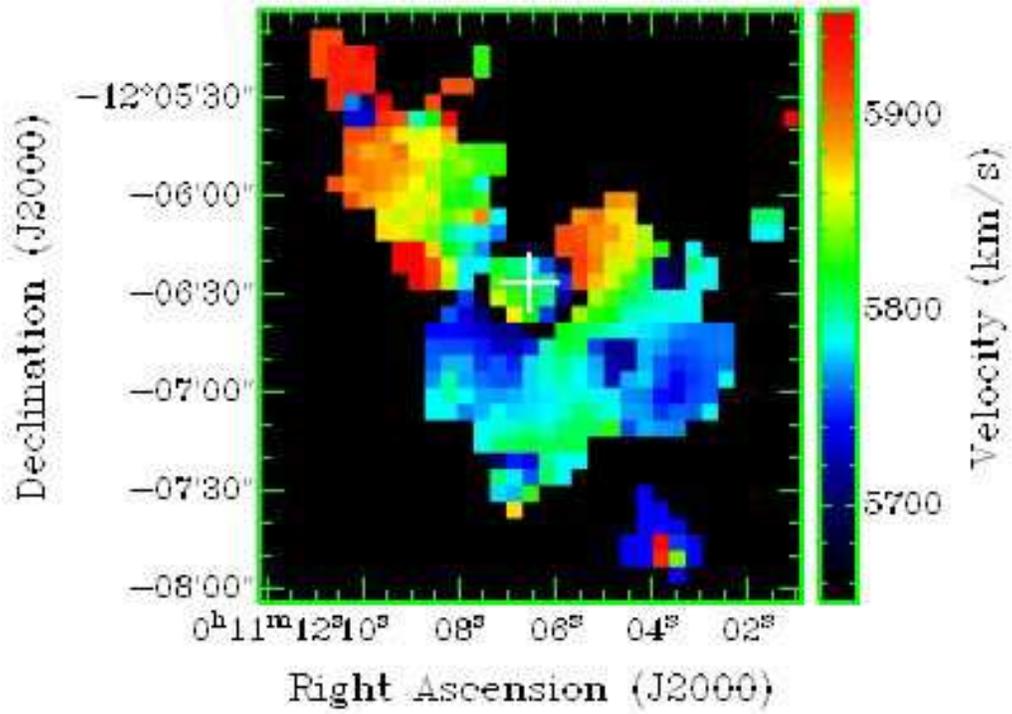}
\vspace{0.5in}
\caption{Moment 1 Map: Color image of the velocity field showing the kinematics of the tidal tails.  The white cross indicates the position of the nucleus.}
\end{center}
\end{figure}

\begin{figure}
\begin{center}
\includegraphics[trim = 15mm 50mm 15mm 50mm,scale=1]{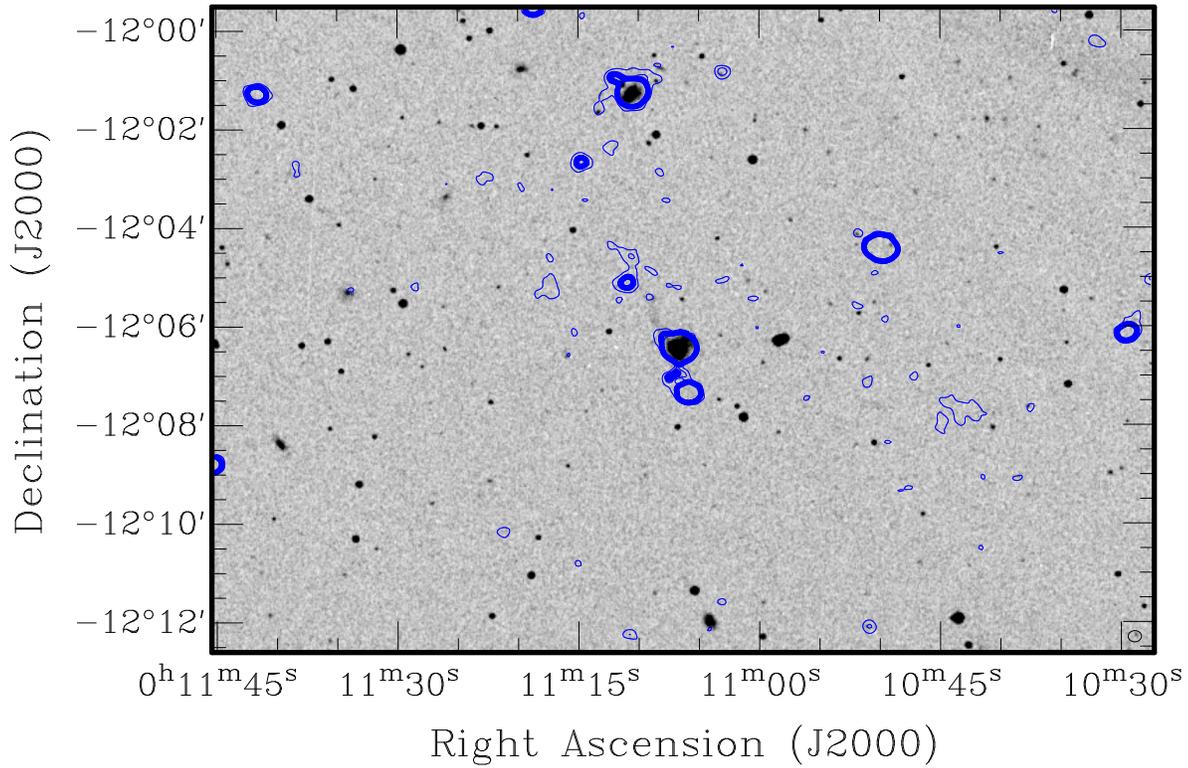}\\
\vspace{0.5in}
\caption{Untapered image of the radio continuum showing nuclear and extra-nuclear emission. Contours are drawn at levels of (0.2, 0.4, 0.6, 0.8, 1) mJy/beam and are shown overlaid on a \textit{Digitized Sky Survey} image.  The size of the synthesized beam is shown in the bottom right hand corner.}
\end{center}
\end{figure}

\begin{figure}
\begin{center}
$\begin{array}{c}
\ST \includegraphics[trim = 15mm 1mm 15mm 25mm,scale=0.5]{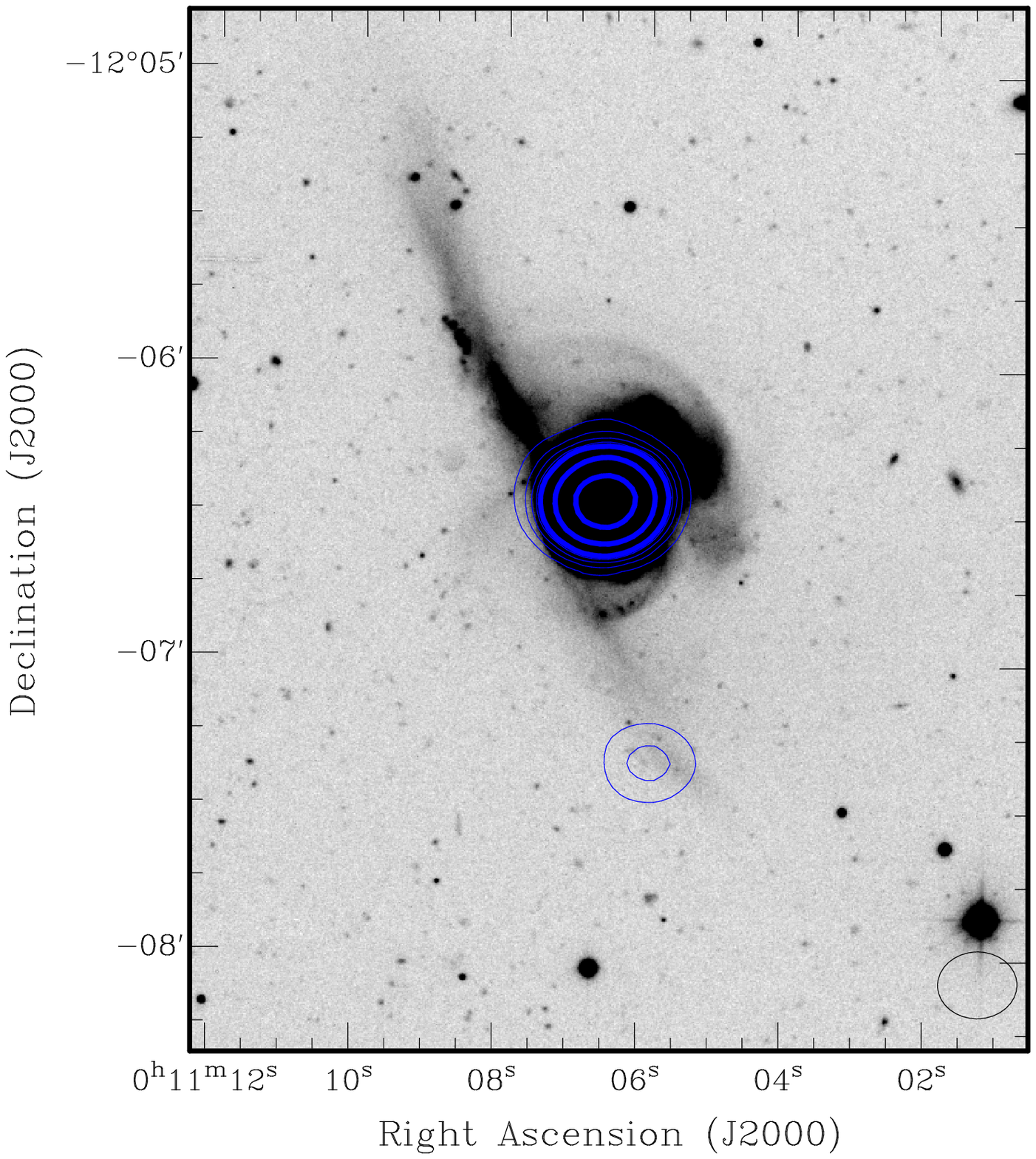}\\
\ST \includegraphics[trim = 15mm 60mm 15mm 70mm, clip,scale=1]{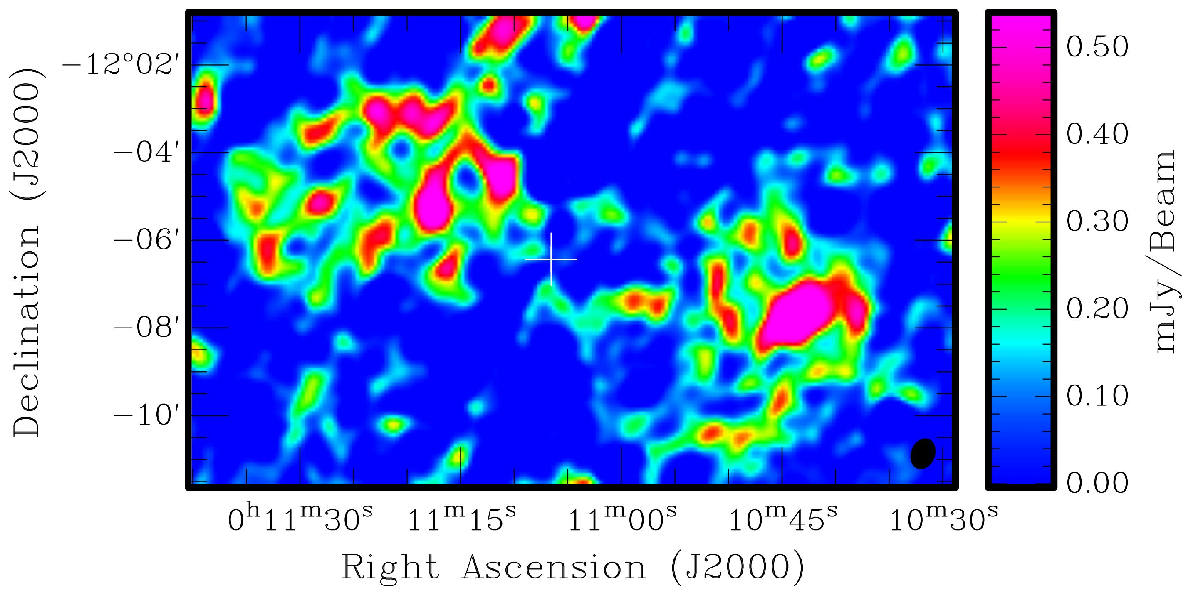}\\
\end{array}$
\vspace{0.1in}
\caption{Radio Continuum Images with the size of the synthesized beam shown in the bottom right hand corner. \textit{Top:} Closer look at the central emission. Contours are drawn at levels of (2, 4, 6, 8, 10, 20, 40) mJy/beam and are shown overlaid on a deep optical image of SS07. \textit{Bottom:} Tapered image of the extended emission without the point sources. The extent of the faint radio lobes is about 390 kpc.}
\end{center}
\end{figure}

\begin{figure}
\begin{center}
\includegraphics[trim = 10mm 10mm 10mm 10mm,scale=1]{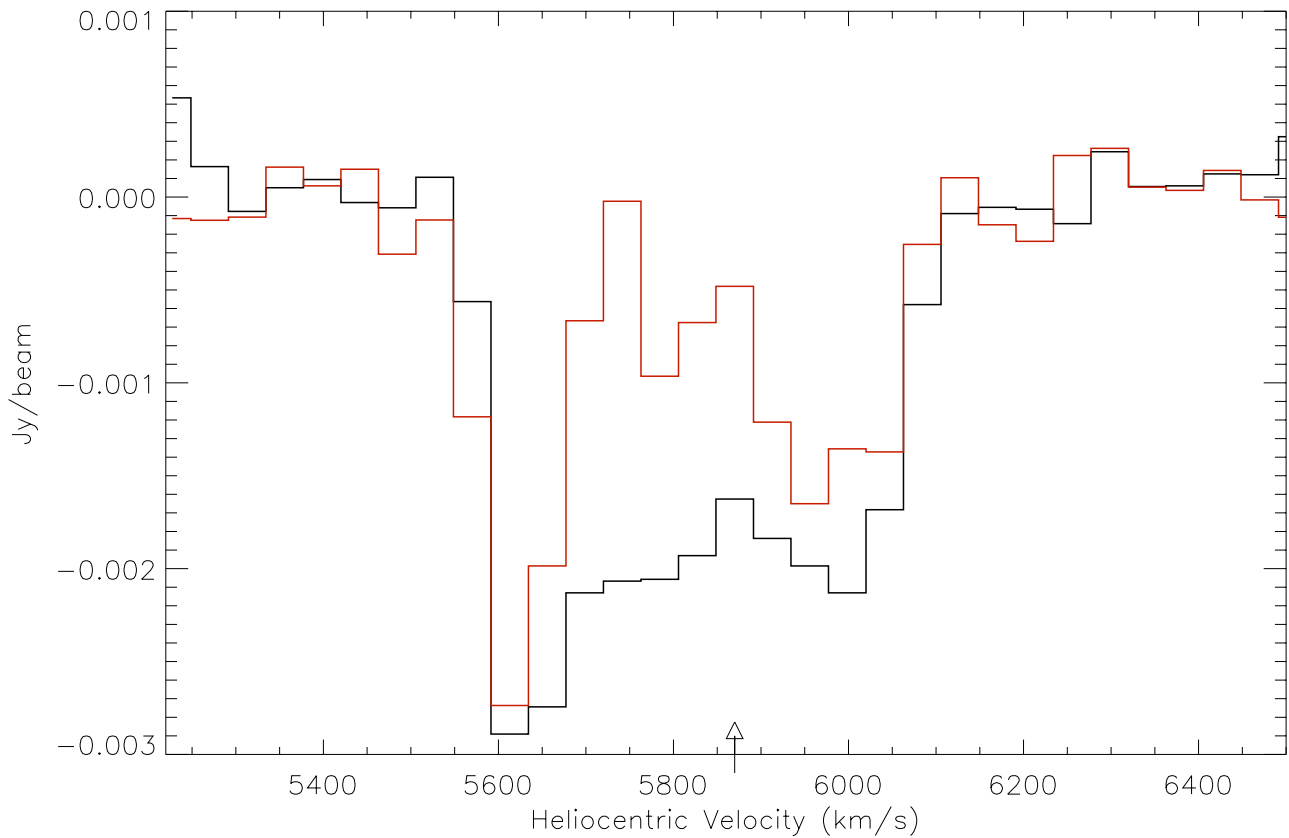}
\vspace{1in}
\caption{Comparison of the absorption profiles observed with the two array configurations (\textit{Black:} CnB Array, \textit{Red:} DnC Array).  Notice that the depth of the two absorption features near 5600 km s$^{-1}$ and 6000 km s$^{-1}$ is similar for the two arrays, while the CnB profile shows significantly deeper absorption around 5800 km s$^{-1}$. This indicates that the DnC array may see emission centered near the systemic velocity.}
\end{center}
\end{figure}

\begin{figure}
\begin{center}
\includegraphics[trim = 10mm 20mm 10mm 10mm,scale=0.85]{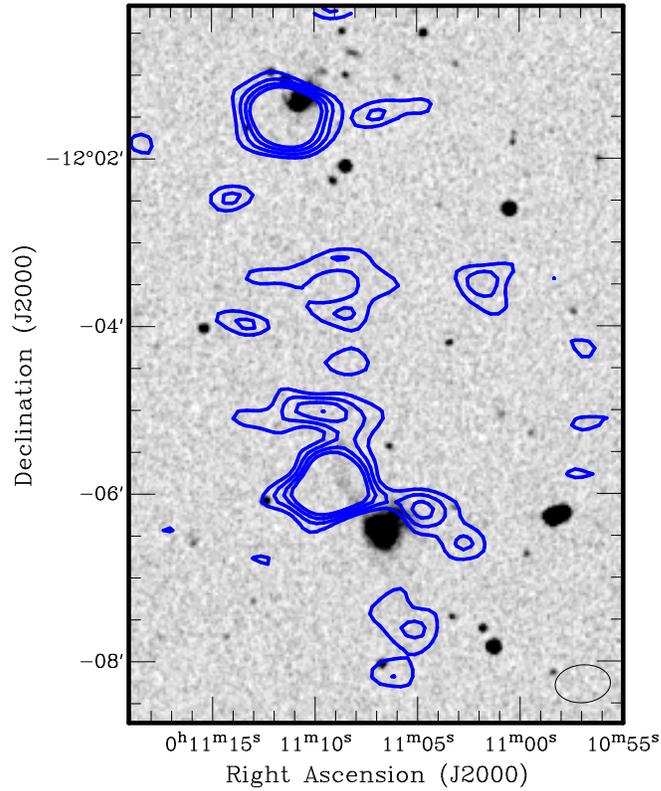}
\vspace{0.5in}
\caption{Channel map of the DnC-array observations at the systemic velocity, overlaid on a \textit{Digitized Sky Survey} image.  The contours are drawn at levels of (0.44, 0.66, 0.88, 1.10) mJy/beam. The size of the synthesized beam is shown in the bottom right hand corner.  This map shows a hint of emission between NGC 34 and NGC 35, possibly indicating a relatively close encounter in the recent past.}
\end{center}
\end{figure}

\begin{figure}
\begin{center}
\includegraphics[trim = 20mm 10mm 10mm 10mm,scale=1]{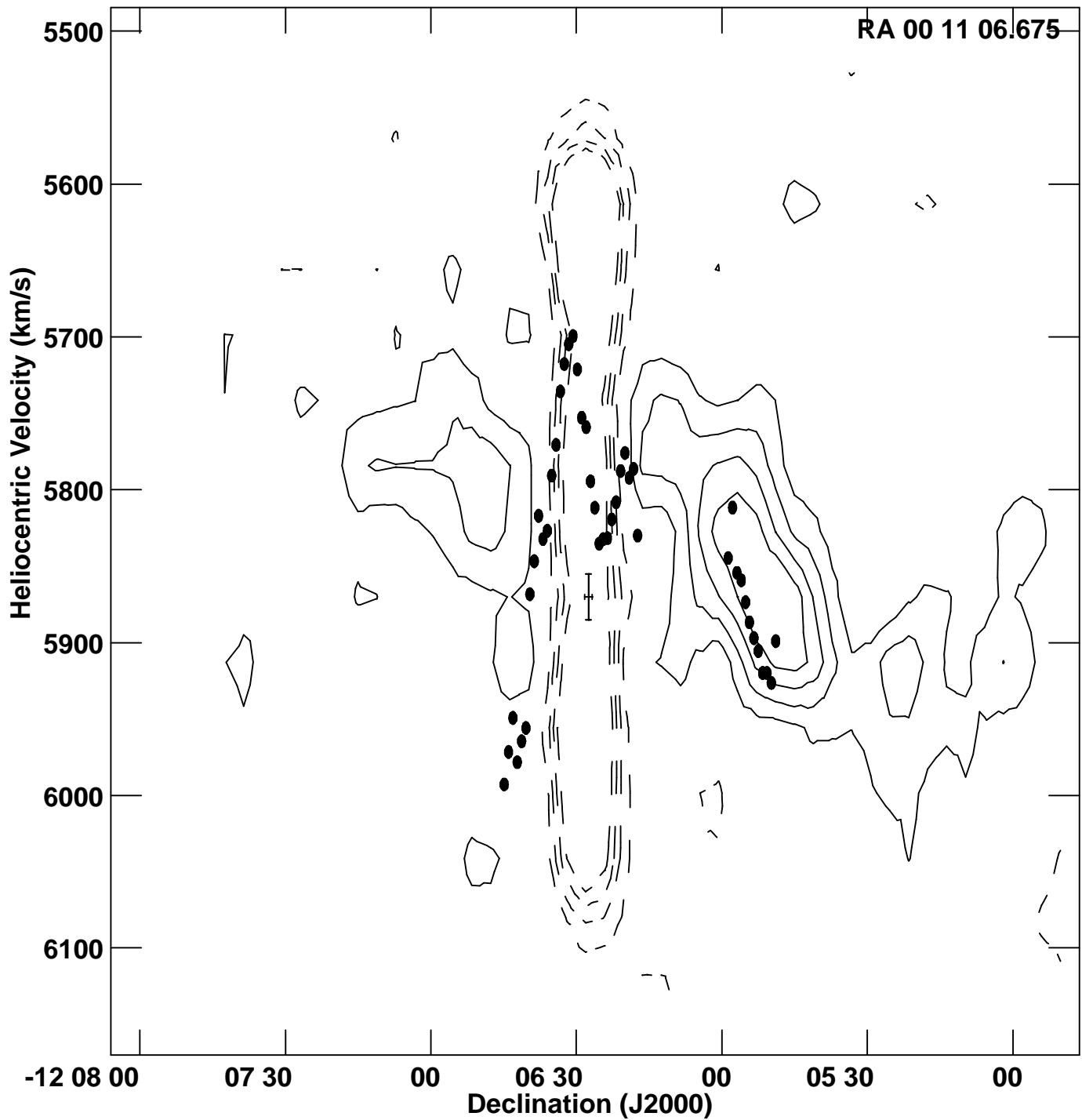}
\caption{Position--velocity diagram along the northern tail and through the gas southwest of the nucleus (same as Fig.\ 3), overlaid with ionized-gas velocities measured at the same position angle from an optical spectrum (data points). The solid lines represent \HI emission, while the dashed ones show the \HI absorption. The cross marks the central declination and optical systemic velocity.}
\end{center}
\end{figure}

\begin{figure}
\begin{center}
\includegraphics[scale=1]{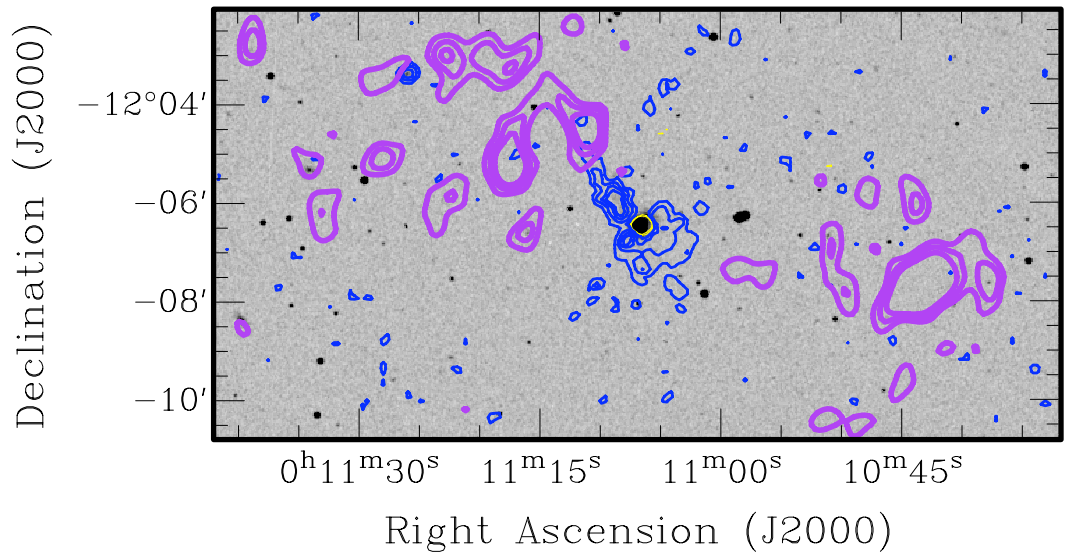}
\vspace{0.5in}
\caption{\HI and radio-continuum observations overlaid on a \textit{Digitized Sky Survey} image. Blue contours represent the \HI emission, while yellow contours represent \HI absorption. Purple contours mark regions of radio-continuum emission (same parameters as Fig.\ 8b). Even though the resolutions are different, this figure serves to compare the spatial extent of the HI and radio-continuum.}
\end{center}
\end{figure}

\clearpage

\begin{deluxetable}{cccccc}
\tabletypesize{\small}
\tablenum{1}
\tablecolumns{6}
\tablewidth{0pc}
\tablecaption{Observation Parameters}
\tablehead{
\colhead{}&
\colhead{DnC\tablenotemark{1}}& 
\colhead{CnB\tablenotemark{2}}& 
\colhead{DnC+CnB} & 
\colhead{Untapered}&
\colhead{Tapered}\\
\colhead{}&
\colhead{}&
\colhead{}&
\colhead{}&
\colhead{Continuum}&
\colhead{Continuum}}
\startdata
  Number of Channels & 31 & 31 & 31 &  11 & 11 \\
  Flux Calibrator & 0137+331 & 0137+331 & - & -& -\\
  Phase Calibrator & 0018-127 & 2357-114 & - & - & - \\
  Synthesized Beam &  $39.6^{\prime\prime}\times27.1^{\prime\prime}$ & $14.2^{\prime\prime}\times12.0^{\prime\prime}$ & $16.3^{\prime\prime}\times13.6^{\prime\prime}$  & $16.4^{\prime\prime}\times13.6^{\prime\prime}$& $42.1^{\prime\prime} \times31.6^{\prime\prime}$ \\
  Noise (mJy/beam) & 0.22 & 0.14 &0.13 & 0.05 & 0.10 \\
  $T_b$ conversion factor (K) & 0.56 & 3.52 & 2.70 & - & - \\ 
  $N_{\rm HI}$ Sensitivity (cm$^{-2}$) &$ 9.7  \times 10^{18}$ &$ 3.9 \times 10^{19}$ &$ 2.8 \times 10^{19}$ & - & -  \\
 \enddata
\tablenotetext{1}{Observed in June 2008}
\tablenotetext{2}{Observed in June 2009}
\end{deluxetable}

\begin{deluxetable}{cccccc}
\tablenum{2}
\tablecolumns{5}
\tablewidth{0pc}
\tablecaption{Gas-Rich Group of Galaxies}
\tablehead{
\colhead{Name}&
\colhead{R.~A.}& 
\colhead{Decl.}&
\colhead{$V_{\rm HI,\,hel}$}&
\colhead{$\int{Sdv}$}&
\colhead{$M_{\rm HI}$}\\
\colhead{}&
\colhead{(J2000)}&
\colhead{(J2000)}&
\colhead{(km $\rm s^{-1}$)}&
\colhead{(Jy km $\rm s^{-1}$)}&
\colhead{( $10^9 ~ \rm M_ {\odot}$)}}
\startdata
NGC 34      & 00 11 06.56 & $-$12 06 27.5 & ...\tablenotemark{1} & $4.20 \pm 0.13$ & $7.20 \pm 0.22$\\ 
NGC 35      & 00 11 10.50 & $-$12 01 15.2 & 5977 & $4.26 \pm 0.10$ & $7.30 \pm 0.17$ \\
PGC 958866  & 00 11 19.68 & $-$12 00 44.4 & 6021 & $0.61 \pm 0.05$ & $1.04 \pm 0.09$ \\
PGC 958282  & 00 11 25.91 & $-$12 03 20.4 & 5763 & $0.16 \pm 0.02$ & $0.27 \pm 0.03$ \\
Uncatalogued& 00 10 30.54 & $-$12 02 11.2 & 5978 & $0.58 \pm 0.05$ & $0.99 \pm 0.09$ \\
\enddata
\tablenotetext{1}{Disturbed and complex integrated \HI profile yields no reliable value for the systemic velocity.}
\end{deluxetable}

\end{document}